\documentclass{article}

\usepackage{arxiv}

\usepackage[utf8]{inputenc} % allow utf-8 input
\usepackage[T1]{fontenc}    % use 8-bit T1 fonts
\usepackage{hyperref}       % hyperlinks
\usepackage{url}            % simple URL typesetting
\usepackage{booktabs}       % professional-quality tables
\usepackage{amsfonts}       % blackboard math symbols
\usepackage{nicefrac}       % compact symbols for 1/2, etc.
\usepackage{microtype}      % microtypography
\usepackage{lipsum}		% Can be removed after putting your text content
\usepackage{graphicx}
\usepackage{natbib}
\usepackage{doi}

\title{Style Augmentation improves Medical Image Segmentation}

\author{ \href{https://orcid.org/0000-0000-0000-0000}{\includegraphics[scale=0.06]{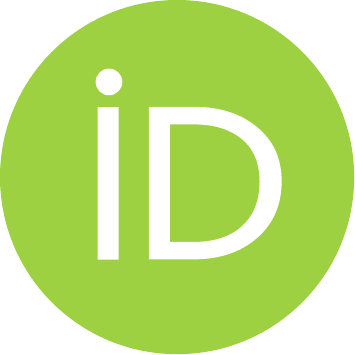}\hspace{1mm}Kevin Ginsburger} \\
	Commissariat à l'Energie Atomique,\\
	Chemin du Ru,\\
	91680 Bruyères-le-Châtel \\
	\texttt{Kevin.GINSBURGER2@cea.fr} \\
}

\renewcommand{\headeright}{Technical Report}
\renewcommand{\undertitle}{Technical Report}
\renewcommand{\shorttitle}{Style Augmentation improves Medical Image Segmentation}

\hypersetup{
pdftitle={Style Augmentation improves Medical Image Segmentation},
pdfsubject={cs.CVPR},
pdfauthor={Kevin Ginsburger},
pdfkeywords={Data Augmentation, Deep Learning, Segmentation, CNN Texture Bias},
}

\begin{document}
\maketitle

\begin{abstract}
Due to the limitation of available labelled data, medical image segmentation is a challenging task for deep learning.
Traditional data augmentation techniques have been shown to improve segmentation network performances by optimizing the usage of few training examples. However, current augmentation approaches for segmentation do not tackle the strong texture bias of convolutional neural networks, observed in several studies. This work shows on two medical datasets that style augmentation, which is already used in classification tasks, helps reducing texture over-fitting and improves segmentation performance.  

\end{abstract}


\section{Introduction}

Among the deep learning literature, medical image segmentation has emerged as a specific research topic, owing to the typical size of the available training datasets. Bridging the gap between few-shot and data-heavy learning-based segmentation methods, the infamous U-Net architecture~\citep{ronneberger_u-net_2015} has founded a new family of convolutional neural networks (CNNs) with relative data frugality (typically between a few dozens and a few thousands training examples).
In this data regime, data augmentation plays a major role to prevent over-fitting and improve generalization capabilities of the CNNs. As such, a plethora of augmentation techniques have been designed, using both simple geometric transformations such as rotations, reflections and elastic deformations as well as more advanced approaches. Most of these techniques modify the shape, color or contrast of the images while preserving the texture~\citep{shorten_survey_2019}.
However, recent works strongly suggest that CNNs are biased towards texture~\citep{geirhos_imagenet-trained_2019, azad_texture_2020}. Quantitative experiments demonstrated that style augmentation, which randomizes texture, contrast and color, improves classification performance on several benchmark datasets~\citep{jackson_style_2019, geirhos_imagenet-trained_2019}.
Despite these appealing performance boosts in classification tasks, the interest of style augmentation has not yet been evaluated on segmentation tasks. Medical segmentation seems particularly adapted to this kind of experiment. Indeed, small medical training datasets are not always sufficiently representative of the full spectrum of textures the network might encounter during evaluation. As such, changes of imaging devices, acquisition setups, or tissue characteristics might lead to performance drops with a texture-biased network.
In this work, the interest of style augmentation for segmentation tasks is evaluated on the MoNuSeg dataset. Using the recent UNeXt architecture with reduced number of parameters, the conducted experiments show that style augmentation strongly prevents over-fitting and improves segmentation performance.

\section{Methods}
\label{sec:methods}

Studies on stylization and style augmentation applied to classification and regression tasks~\citep{geirhos_imagenet-trained_2019, azad_texture_2020} showed that these approaches are particularly beneficial in cases where labelled data are scarce, and a good shape representation is needed to be able to generalize to the test dataset.
In what follows, similar training conditions are used to evaluate the interest of style augmentation on a segmentation task.

\subsection{Dataset}

The Multi-Organ Nucleus Segmentation (MoNuSeg)~\citep{kumar_dataset_2017, kumar_multi-organ_2020} dataset seems particularly adapted to the experiment at hand. It is both small and diverse, and contains H\&E stained tissue images from multiple organs and patients. It consists of 30 training images (for each experiment: 25 images are used for training and 5 are kept for validation to keep track of over-fitting) and 14 test images, all resized to 512x512 for the experiments.

\subsection{Segmentation network}

The state-of-the-art UNeXt architecture is employed, with a 10\% dropout rate, a mini-batch size of $4$ and $2000$ training epochs. The model with best Intersection over Union (IoU) score on the validation set is saved during training. Dropout is also applied during inference. Segmentation performance metrics are thus averaged over 20 instances of the network, on which a 10\% dropout rate is applied. Combining early convolutional stages with MLPs in the latent space, UNeXt significantly reduces the number of parameters while boosting segmentation performance. This minimal complexity is of particular interest for small datasets which are prone to over-fitting. It also enables to conduct multiple experiments efficiently.

\subsection{Data augmentation}

\begin{figure}
	\centering
	\includegraphics[width=.4\linewidth]{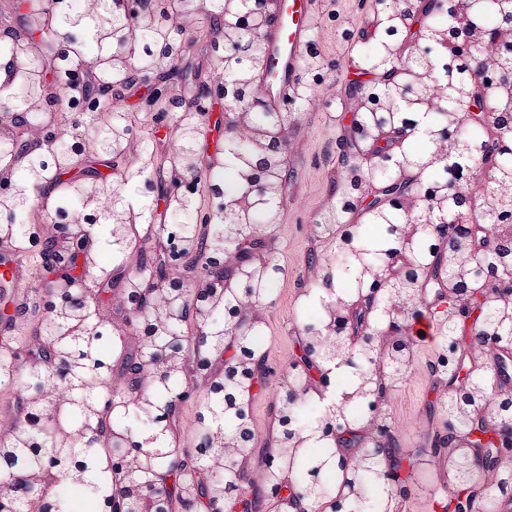}
	\includegraphics[width=.4\linewidth]{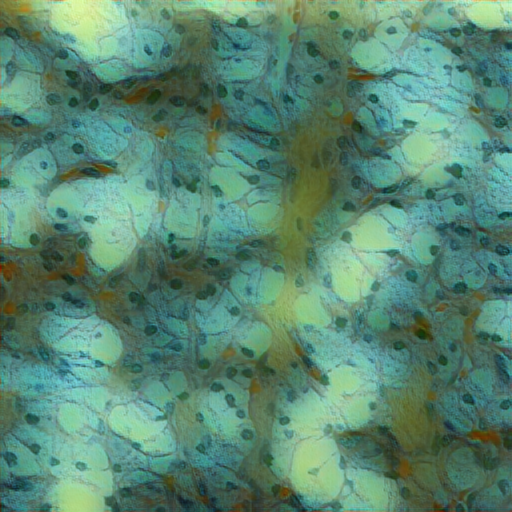}
	\caption{Example training image (left) and corresponding image with style augmentation applied (right).}
	\label{fig:example_style}
\end{figure}

Traditional augmentation techniques are used in all conducted experiments, combining random 90 degrees rotations and flipping.
The style augmentation implementation follows the method proposed in~\citep{jackson_style_2019}, with a style transfer strength set to the optimal value of $0.5$. This approach offers great inference speed and enables to apply a broad range of random styles while preserving shapes and edges.  
During UNeXt training, a random stylization ratio is drawn for each mini-batch, which sets the number of training examples with randomized styles for this batch. Example stylized images are shown in figure \ref{fig:example_style}.

\section{Results}
\label{sec:results}

\begin{figure}
	\centering
	\includegraphics[width=.4\linewidth]{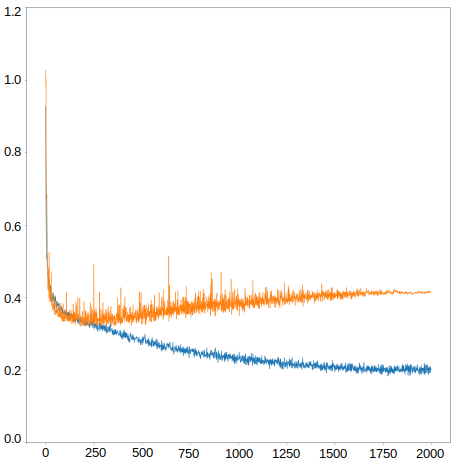}
	\includegraphics[width=.4\linewidth]{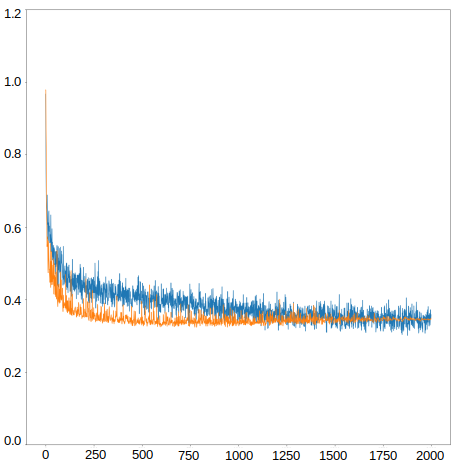}
	\caption{Training (blue) and validation (orange) losses against epochs without style augmentation (left) and with style augmentation (right).}
	\label{fig:losses}
\end{figure}

Figure~\ref{fig:losses} shows the train and validation losses (blue and orange curves respectively) obtained during the training of UNeXt on the MoNuSeg dataset. The left graph correspond to the basic training scheme where only flipping and random 90 degrees rotations are employed for augmentation, while style augmentation was added in the right graph, following the method described in the previous section.

\begin{figure}
	\centering
	\includegraphics[width=.4\linewidth]{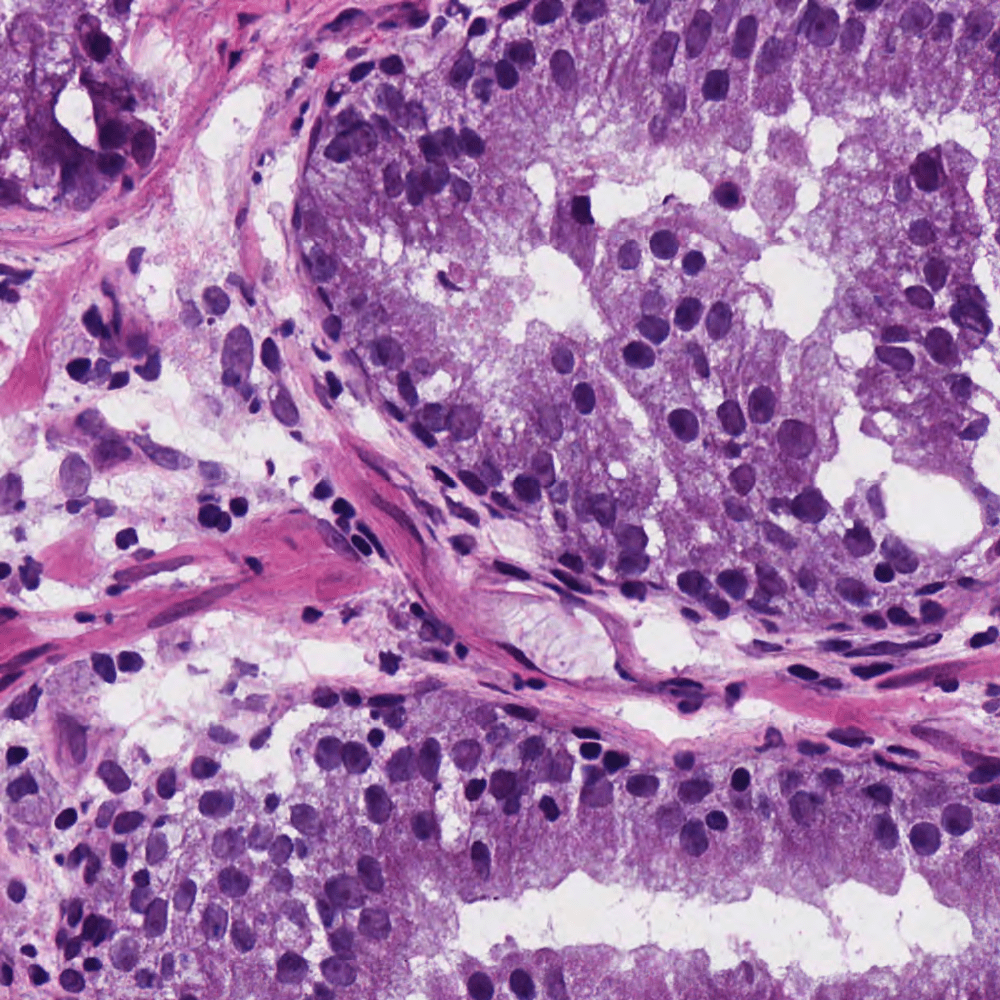}
	\includegraphics[width=.4\linewidth]{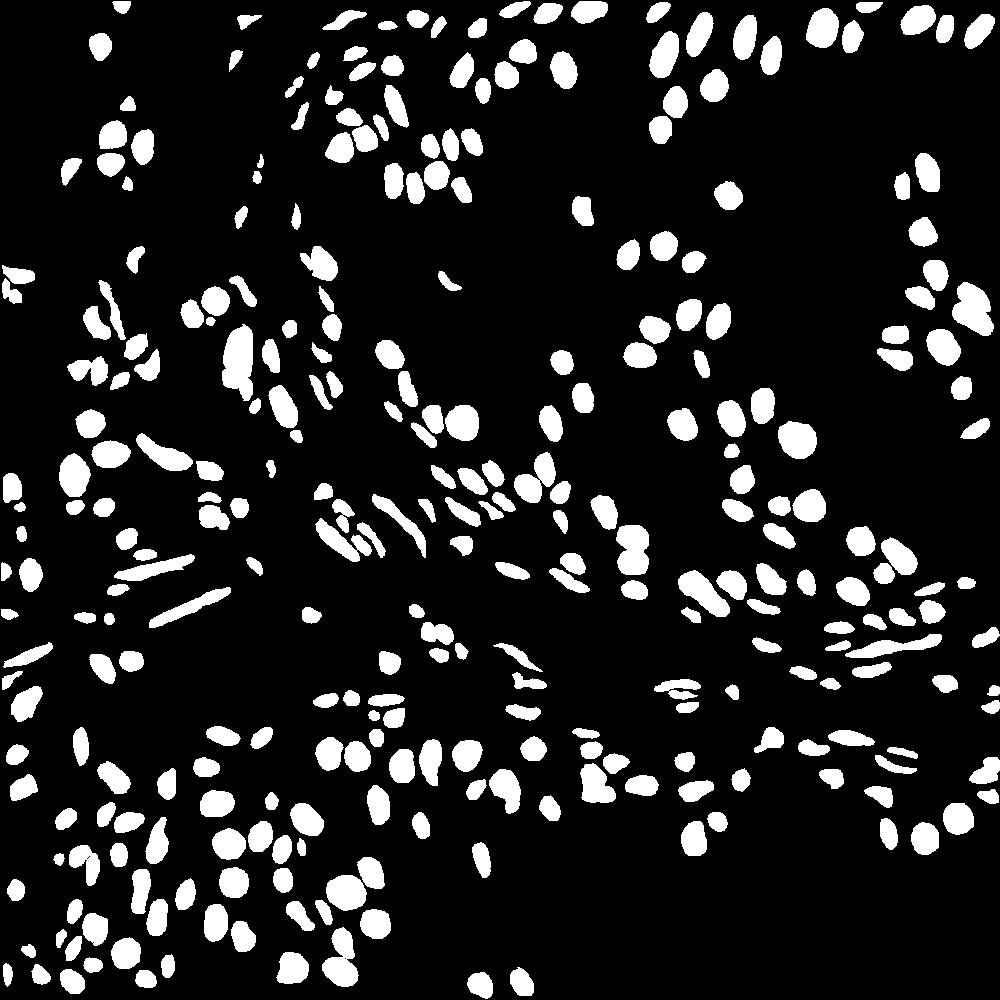}
	\includegraphics[width=.4\linewidth]{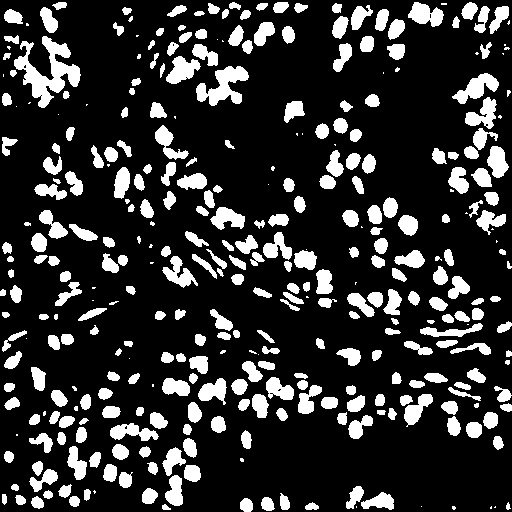}
	\includegraphics[width=.4\linewidth]{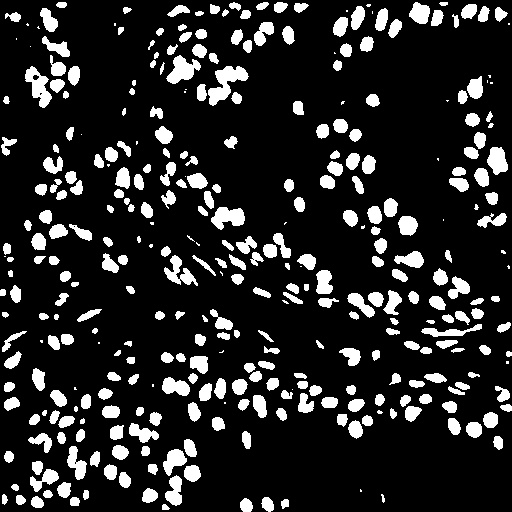}
	\caption{Example test image (upper left), corresponding ground thruth segmentation (upper right), UNeXt segmentation without style augmentation (lower left), UNeXt segmentation with style augmentation (lower right).}
	\label{fig:quali_results}
\end{figure}

The effect of style augmentation is qualitatively illustrated in figure~\ref{fig:quali_results}.
The original image and ground truth segmentation from the MoNuSeg test set are shown on the top. Bottom row compares binary segmentation masks obtained from the UNeXt network trained without and with style augmentation (left and right respectively).

The quantitative effect of style augmentation on customary segmentation metrics is given in table~\ref{tab:quanti_results}. 
Both Dice score and Intersection over Union (IoU) are computed for the two UNeXt networks trained without and with style augmentation.

\begin{table}
	\caption{Intersection over Union (IoU) and Dice score obtained on the MoNuSeg dataset with UNeXT. No Style Aug./Style Aug.: metrics without/with style augmentation used during training of UNeXt.}
	\centering
	\begin{tabular}{ccc}
		\cmidrule{2-3}
		~ & No Style Aug. & Style Aug. \\
		\midrule
		IoU & $0.6072$ & $0.6656$ \\
		\midrule
		Dice & $0.7533$ & $0.7991$ \\
		\bottomrule
	\end{tabular}
	\label{tab:quanti_results}
\end{table}

\section{Discussion}
\label{sec:discussion}

The first interest of style augmentation observed in the conducted experiment is the prevention of over-fitting. As shown in the left graph of figure~\ref{fig:losses}, training UNeXt on the MoNuSeg dataset without style augmentation leads to early over-fitting. The validation loss increases at the 200\emph{th} epoch, while the training loss is still decreasing.
On the contrary, on the right graph, style augmentation leads to a smooth convergence of training and validation losses to the same value. The insertion of new stylized images in each batch makes training more difficult, which results in a noisier loss curve, and a slower decrease of the training loss. The training loss converges to a higher value, but the final the validation loss is smaller.

As outlined in figure~\ref{fig:quali_results}, the good training properties of style augmentation reflect qualitatively on the obtained segmentation masks during the test phase. The segmentation mask obtained from the basic training strategy exhibits more false positives than the one obtained with style augmentation. For instance, in the right bottom of the image, style augmentation gives a segmentation masks quite close to the ground truth, while the basic approach segments cells erroneously. Style augmentation also yields more precise segmentation masks, with sharper edges and less errors around segmented cells.

Table~\ref{tab:quanti_results} also indicates quantitative improvements on IoU and Dice scores, with a significant progress on both metrics using style augmentation. Moreover, while only 25 of the 30 images of the dataset were used for each training (5 being kept for validation), UNeXt trained with style augmentation matches state-of-the-art quantitative results on the MoNuSeg dataset obtained with the Medical Transformer setup~\citep{valanarasu_medical_2021}.

\section{Conclusion}
\label{sec:conclusion}

Adding random styles to training images is an easily implemented augmentation technique, with low computational overhead, which can be readily applied to any training scheme without adding hyperparameters.
Preliminary results presented in this study on the UNeXt segmentation network trained on the MoNuSeg dataset indicate a clear benefit of style augmentation to prevent over-fitting and boost segmentation performance.  
While further experiments are needed to validate the efficiency of the proposed approach on other datasets and segmentation networks, these first results demonstrate the interest of style augmentation for learning-based segmentation on a small and heterogeneous dataset.
For segmentation tasks in the low data regime, style augmentation could be advantageously combined to image generation techniques. Indeed,~\citep{thambawita_singan-seg_2022} demonstrated the ability of SinGAN to produce synthetic training pairs with realistic shapes. Adding random styles to the synthesized images could alleviate texture over-fitting and favor a shape-based representation for segmentation.

\bibliographystyle{unsrtnat}
\bibliography{style_aug}

\end{document}